\documentclass[10pt]{iopart}

\usepackage{iopams}  
\usepackage{graphicx}
\usepackage{amssymb}
\usepackage{epstopdf}
\usepackage{bbold}
\usepackage{float}
\usepackage{color}
\usepackage{subfigure}
\usepackage{cite}
\usepackage{relsize}
\expandafter\let\csname equation*\endcsname\relax

\expandafter\let\csname endequation*\endcsname\relax

\usepackage{mathtools}
\usepackage[ngerman, english]{babel}

\ioptwocol

\begin{document}

\title{Triatomic butterfly molecules}

\author{Matthew T. Eiles$^{1*}$, Christian Fey$^2$, Frederic Hummel$^{3}$, Peter Schmelcher$^{3,4}$}

\address{$^1$Max-Planck-Institut f\"ur Physik komplexer Systeme, N\"othnitzer Str.\ 38,
01187 Dresden, Germany\newline
$^2$ Max-Planck-Institute of Quantum Optics, 85748 Garching, Germany\newline
$^3$Zentrum f\"ur Optische Quantentechnologien, Fachbereich Physik, Universit\"at Hamburg, Luruper Chaussee 149, 22761 Hamburg, Germany\newline
$^4$The Hamburg Centre for Ultrafast Imaging, Universit\"at Hamburg, Luruper Chaussee 149, 22761 Hamburg, Germany}

\vspace{10pt}
\ead{\mailto{meiles@mpg.pks.de*}}
\begin{abstract}
We detail the rich electronic and vibrational structure of triatomic ``butterfly'' molecules, ultra-long-range Rydberg molecules bound by resonant $p$-wave scattering.  We divide these molecules into two sub-classes depending on their parity under reflection of the electronic wave function through the molecular plane. The trimers with odd reflection parity have topographically smooth potential energy surfaces except near the collinear configuration. Here, the vibrational wave function is confined tightly in the symmetric-stretch and bending modes, but only loosely in the asymmetric stretch mode. The trimers with even reflection parity exhibit far richer potential surfaces with abundant minima, but only a few of these are deep enough to localize the vibrational states. These minima are correlated with the electronic wave functions of the butterfly dimer, contributing to a building principle for trimers.

\end{abstract}

%

%
%
%

%

\newcommand{\be}{\begin{equation}}
\newcommand{\ee}{\end{equation}}
\newcommand{\lap}{\nabla^2}
\newcommand{\pd}[1]{\frac{\partial}{\partial{#1}}}
\newcommand{\pdd}[1]{\frac{\partial^2}{\partial{#1}^2}}
\newcommand{\pdde}[2]{\frac{\partial^2{#1}}{\partial{#2}^2}}
\newcommand{\pde}[2]{\frac{\partial{#1}}{\partial{#2}}}
\newcommand{\ar}{(\vec{r},t)}
\newcommand{\arz}{(\vec{r},0)}
\newcommand{\psirt}{\psi(\vec{r},t)}
\newcommand{\psirz}{\psi(\vec{r},0)}
\newcommand{\psirzc}{\psi^*(\vec{r},0)}
\newcommand{\psirtc}{\psi^*(\vec{r},t)}
\newcommand{\probcur}{\vec{S}\ar}
\newcommand{\twodvec}[2]{\left[\begin{array}{cc}{#1} \\ {#2}\end{array}\right]}
\newcommand{\twodmat}[4]{\left[\begin{array}{cc}{#1} & {#2} \\ {#3} & {#4}\end{array}\right]}
\newcommand{\threedvec}[3]{\left[\begin{array}{ccc}{#1} \\ {#2} \\ {#3}\end{array}\right]}
\newcommand{\threedmat}[9]{\left[\begin{array}{ccc}{#1} & {#2}& {#3} \\ {#4} & {#5} & {#6}\\ {#7} & {#8} & {#9}\\\end{array}\right]}
\newcommand{\fourdvec}[4]{\left[\begin{array}{cccc}{#1} \\ {#2} \\ {#3} \\ {#4}\\ \end{array}\right]}
\newcommand{\fourdmat}[4]{\left[\begin{array}{cccc}{#1}\\{#2}\\{#3}\\{#4}\\\end{array}\right]}
\newcommand{\bra}[1]{\langle{#1}|}
\newcommand{\ket}[1]{|{#1}\rangle}
\newcommand{\bkt}[2]{\langle{#1}|{#2}\rangle}
\newcommand{\apdag}{a_+^\dagger}
\newcommand{\amdag}{a_-^\dagger}
\newcommand{\am}{a_-}
\newcommand{\ap}{a_+}
\newcommand{\sol}{\textbf{Solution: }}
\newcommand{\sumn}{ \sum_{n = 0}^N}
\newcommand{\com}[2]{\left[{#1},{#2}\right]}
\newcommand{\h}{\frac{1}{2}}
\newcommand{\hh}{\frac{1}{2}}
\newcommand{\dd}[1]{\mathrm{d}{#1}}
\newcommand{\ddd}[1]{\mathrm{d}^3{#1}}
\newcommand{\ddn}[2]{\mathrm{d^{#1}}{#2}}
\newcommand{\up}{\ket{\uparrow}}
\newcommand{\dn}{\ket{\downarrow}}
\newcommand{\upb}{\bra{\uparrow}}
\newcommand{\dnb}{\bra{\downarrow}}
\newcommand{\vv}[1]{\underline{#1}}
\newcommand{\bigO}{\mathcal{O}}
\newcommand{\del}{\underline \nabla}
\newcommand{\+}[1]{{#1}}
\newcommand*\rfrac[2]{{}^{#1}\!/_{#2}}
\newcommand{\Bohr}{\,\textrm{a}_0}
\def\matt{\textcolor{blue}}

\section{Introduction}
With very few exceptions, atomic negative ions -- weakly bound systems composed of an electron ($e^-$) and a neutral atom ($B$) -- possess only a single bound state \cite{Andersen,BuckmanClark}. In the alkali atoms this is the $^1S$ state, bound by about 500meV.  For several years  the existence of an excited $^3P$ state in cesium was under debate until photodetachment experiments eventually revealed that it is in fact an unbound shape resonance \cite{ChrisJJ,pwaveresonance2,pwaveresonance3,pwaveresonance1,BahrimThumm,BahrimThumm2,FabrikantNegIons}. Indeed, all alkali species possess a $p$-wave shape resonance just a few meV above threshold \cite{Andersen,BuckmanClark, EilesHetero}.

Despite the transient nature of these resonances, they are responsible for the formation of a  class of ultra-long-range Rydberg molecules known as ``butterflies.'' A butterfly molecule consists of a Rydberg atom $(B^+ + e^-)$ bound to a neutral ground state atom ($B$) via the $e^-+B$ $p$-wave scattering interaction, which leads at resonance to a short-lived ion-pair state  \cite{HamiltonGreeneSadeghpour,KhuskivadzeJPB,KhuskivadzePRA}. Superimposed onto this ion-pair potential is the oscillatory structure of the Rydberg wave function, and vibrational states form in the resulting potential wells. Butterfly molecules have been observed in rubidium and, due to their large dipole moments, exhibit pendular behavior in weak external fields \cite{Butterfly, EilesPendular}.

In this article, we show that this $p$-wave binding mechanism can bind a second ground-state atom to the Rydberg atom, forming a triatomic butterfly molecule. Other ultra-long-range triatomic Rydberg molecules formed by the $s$-wave scattering interaction have been studied previously theoretically and experimentally \cite{Rost2006,Rost2009,quantumreflection,JPBdens,FeyKurz,EilesHyd,FeyNew,FeyTrimer}, but a study of the butterfly trimer's full electronic and vibrational structure has not yet been attempted. We determine this structure by computing, analyzing, and interpreting the underlying three-dimensional potential energy surfaces based on the Born-Oppenheimer approximation and the resulting nuclear eigenstates. The latter are obtained using a combined discrete variable and finite difference approach. This effort is simplified by the fact that the potential surfaces decouple into two groups distinguished by their electronic parity under reflection through the molecular plane.  We refer to these two classes as \textit{odd} and \textit{even} butterfly trimers. The equilibrium geometries supporting trimer states vary greatly between these two classes: the odd butterflies have only a few minima in a constrained range of possible molecular geometries close to the collinear arrangement, whereas the even butterflies possess a plethora of equilibrium configurations with a rich diversity in the electronic character. The geometries at which these minima occur can be analyzed and understood from features in the electronic wave functions of the diatomic butterfly molecule. 

In section \ref{sec:theory} we discuss our approach to the electronic structure of the trimers and, specifically, their Born-Oppenheimer adiabatic potential energy surfaces. Sections \ref{sec:potentials} and \ref{sec:minima} are dedicated to a discussion of the equilibrium configurations and geometries and the development of a building principles for the even trimers, respectively. Section \ref{sec:states} analyzes the vibrational dynamics of our butterfly trimers. Finally, section \ref{sec:conclusions} contains our brief conclusions and an outlook. 
\section{Computational approach to the electronic structure and Born-Oppenheimer potential energy surfaces}
\label{sec:theory}
\begin{figure}[t]
\includegraphics[width = 0.6\textwidth]{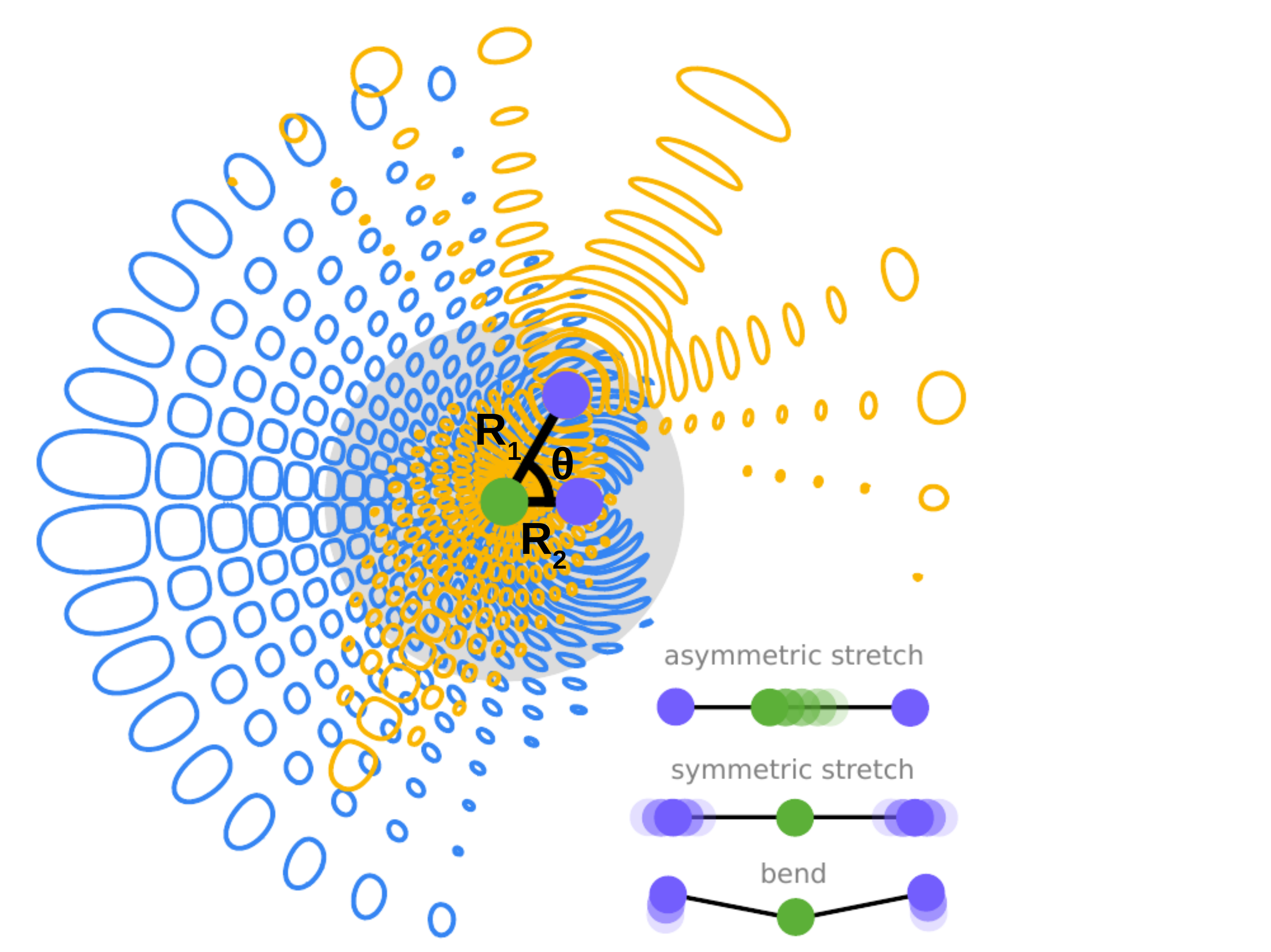}
\caption{A schematic of the $n=30$ butterfly trimer. The two ground-state atoms are marked in purple, and the Rydberg atom is shown in green. The approximate range of bond lengths considered in this butterfly regime lie within the shaded region. The electronic densities of the $\theta_2$-butterfly and the $R_1$-butterfly are depicted in blue and yellow contours, respectively, where each contour specifies when the wave function amplitude evaluated in the molecular plane equals $1\times 10^{-4.8}$. The two bond lengths and bending angle are labeled. On the lower right the three types of nuclear motion exhibited by the trimer are depicted.} 
 \label{introscheme}
\end{figure}

The eigenenergies of the electronic Hamiltonian for fixed nuclei represent the Born-Oppenheimer potential energy surfaces. For our trimers, they depend on the two bond lengths, $R_1$ and $R_2$, and a single  bending angle, $\theta_2=\theta$.  This geometry is illustrated in Fig. \ref{introscheme}. The trimer can exhibit three vibrational modes: a symmetric stretch, in which $R_1=R_2$; an asymmetric stretch, in which $R_1$ grows while $R_2$ shrinks (or vice versa); and a bending mode in which $\theta$ oscillates.  

The interaction of the electron with a ground-state atom is given by the Fermi pseudopotential, generalized by Omont to arbitrary partial waves \cite{Fermi,Omont}.  Including contributions from $s$- and $p$-scattering partial waves only, we obtain the Hamiltonian
\begin{align}
\label{ham}
H({\vec r};R_1,R_2,\theta_{12}) 
& =  -\sum_{n lm}\frac{\ket{n lm}\bra{n lm}}{2(n-\mu_l)^2} \\&+ 2\pi \sum_{i=1}^{2}\sum_{\xi = 1}^4a_i^{(\xi)}\ket{i\xi}\bra{i\xi}.\nonumber
\end{align} 
The first line describes the Rydberg atom using its known eigenfunctions $\phi_{nlm}(\vec r) = \frac{u_{nl}(r)}{r}Y_{lm}(\theta,\varphi) = \bkt{\vec r}{nlm}$ and eigenenergies $-\frac{1}{2(n-\mu_l)^2}$, where $n$ is the principal quantum number and $l$ and $m$ are the orbital and magnetic quantum numbers. For a given $n$ only a few states with $l\le l_\text{min}$ have non-vanishing quantum defects $\mu_l$ which shift them out of the degenerate manifold of high-$l$ states. The second line of Eq. \ref{ham} describes the electron-atom interactions using the Fermi-Omont pseudopotential operator, $\hat V^{(i\xi)} = \ket{i\xi}\bra{i\xi}$, which has the following matrix representation in the Rydberg basis,
\be
\label{fermi}
\hat V^{(i\xi)}_{n'l'm',nlm}=\left. \tilde\partial_\xi\left[\phi_{n' l'm'}(\vec r)\right]^*\tilde\partial_\xi\left[\phi_{n lm}(\vec r)\right]\right|_{\vec r = \vec R_i}.
\ee
This employs a shorthand for the derivative operators used in the  pseudopotentials: $\tilde\partial_{\xi=1}=1$, $\tilde\partial_{\xi=2}=\partial_r$, $\tilde\partial_{\xi=3}=\frac{1}{r}\partial_\theta$, and $\tilde\partial_{\xi=4}=\frac{1}{r\sin\theta}\partial_\varphi$. The three $\xi>1$ terms correspond to the three components of the gradient in the $p$-wave operator. The scattering volumes are  $a_i^{(\xi=1)} = a_s[k(R_i)]$  and $a_i^{(\xi>1)}=3a_p^3[k(R_i)]$.  Eq. \ref{ham} neglects all spin degrees of freedom, and assumes the scattering occurs only in the triplet channel \cite{AndersonPRA, EilesSpin, Hummel,Hummel2}. 

A convenient approach to diagonalize Eq. \ref{ham} has been developed which does not require the complete Rydberg basis. For a given $i$, $\xi$, $n$, and for $l>l_\text{min}$, the matrix defined in Eq. \ref{fermi} has a single non-trivial eigenstate. It follows that there are four ``dimer orbitals'' in total for each atom, $i = 1,2$: one ``trilobite'' for $\xi=1$ , an ``$R_i$-butterfly'' for $\xi=2$, a ``$\theta_i$-butterfly'' for $\xi=3$, and a ``$\varphi_i$''-butterfly for $\xi=4$. Fig. \ref{introscheme} shows two of these dimer orbitals, the $R_1$-butterfly (orange) and $\theta_2$-butterfly (blue). The nodal structures of the butterfly dimer orbitals are arranged such that, at the position of the ground state atom, the wave function changes most rapidly parallel to ($R$-butterfly) or perpendicular to ($\theta$- and $\varphi$-butterflies, in mutually orthogonal directions) the internuclear axis. The $R$-butterfly orbital therefore concentrates electron probability around the internuclear axis, while in the $\theta$- and $\varphi$-butterflies the electronic density fans out over a larger area. The electronic energies of the $\theta$-butterfly and $\varphi$-butterfly dimer orbitals are degenerate. In general, the dimer orbitals are not orthogonal: a $\xi=\alpha$ orbital for the atom located at $\vec R_p$ has an overlap with the $\xi=\beta$ orbital at position $\vec R_q$ equal to \cite{Rost2006,JPBdens,Rost2009,FeyTrimer}
 \be
\label{eqtomdef}
\+{\Upsilon_{pq}^{\alpha\beta}}= \sum_{l>l_\text{min}}^{n-1}\sum_{m=-l}^{m=l}\tilde\partial_\alpha\left[\phi_{nlm}(\vec R_p)\right]^*\tilde\partial_\beta\phi_{nlm}(\vec R_q).
\ee 
Several of these overlap elements vanish at specific geometries. As detailed in Refs. \cite{JPBdens,EilesTutorial}, the projection of the orbital angular momentum of the $\xi=2$ orbital  onto the internuclear axis is zero, while it is unity for the $\xi=3$ and $\xi=4$ orbitals. Therefore, the matrix elements $\Upsilon_{ii}^{23}$ and $\Upsilon_{ii}^{24}$ vanish. Furthermore, the $\xi=4$ orbital has odd parity under reflection through the molecular plane, i.e. $\varphi \to -\varphi$, while the $\xi=2$ orbital is independent of this angle and the $\xi=3$ orbital is an even function of $\varphi$. Thus, both $\xi=2,3$ orbitals have even parity under this same operation. This implies that these orbitals decouple completely with $\xi=4$, i.e. $\Upsilon_{ii'}^{4\beta} = 0$ for $\beta=2,3$. For this reason, in this paper we distinguish between the \textit{even} trimers, which are linear combinations of the $R_1$, $R_2$, $\theta_1$, and $\theta_2$ dimer orbitals, and the \textit{odd} trimers, linear combinations of the two $\varphi_1$ and $\varphi_2$ dimer orbitals. 

Due to the effects of the $p$-wave resonance, the trilobite states are energetically decoupled from the butterfly states and from other $n$ manifolds, and to a good approximation can be studied independently. Refs. \cite{Rost2009,FeyTrimer} have already investigated the trilobite trimer, i.e. the states defined by the Hamiltonian in the $2\times2$ trilobite subspace:
\be
\label{triloeq}
\widetilde{ H}_\text{t} = \begin{pmatrix} a_1^{(1)}[k(R_1)]\Upsilon_{11}^{11} & a_1^{(1)}[k(R_1)]\Upsilon_{12}^{11}\\ a_1^{(1)}[k(R_2)] \Upsilon_{21}^{11} &  a_1^{(1)}[k(R_2)]\Upsilon_{22}^{11}\end{pmatrix},
\ee
where the overall Rydberg energy $-\frac{1}{2n^2}$ has been set to zero. 
This separation of trilobite and butterfly states implies that the butterfly states of a single $n$ manifold are governed by the Hamiltonian $\widetilde{H}_\text{b}$:
\begin{align}
\label{matrix}
\begin{pmatrix}
a_1^{(2)}\+{\Upsilon}_{11}^{22} & a_1^{(2)}\+{\Upsilon}_{21}^{22} & 0 & a_1^{(2)}\+{\Upsilon}_{21}^{32} & 0 & 0 \\
a_2^{(2)}\+{\Upsilon}_{12}^{22} &a_2^{(2)}\+{\Upsilon}_{22}^{22} & a_2^{(2)} \+{\Upsilon}_{12}^{32} & 0 & 0 & 0 \\
0 & a_1^{(3)}\+{\Upsilon}_{21}^{23} &a_1^{(3)} \+{\Upsilon}_{11}^{33} & a_1^{(3)}\+{\Upsilon}_{21}^{33} & 0 & 0\\
a_2^{(3)}\+{\Upsilon}_{12}^{23} &0& a_2^{(3)} \+{\Upsilon}_{12}^{33} & a_2^{(3)}\+{\Upsilon}_{22}^{33} & 0 & 0\\
0 & 0 & 0 & 0 & a_1^{(4)}\+{\Upsilon}_{11}^{44}&  a_1^{(4)}\+{\Upsilon}_{21}^{4 4}\\
0 & 0 & 0 & 0 &  a_2^{(4)}\+{\Upsilon}_{12}^{4 4}& a_2^{(4)}\+{\Upsilon}_{22}^{4 4}\\
\end{pmatrix}.
\end{align}
Note that in Eqs. \ref{triloeq} and \ref{matrix} the Hamiltonian marked by a tilde incorporates is modified from the original Hamiltonian $H$ since it includes the effect of the non-orthogonal basis:
\be\underbrace{\Upsilon^{-1}H}_{\widetilde{H}}\psi = E\underbrace{\Upsilon^{-1}\Upsilon}_{\mathbb{1}}\psi.\ee

The studies of trilobite trimers in Refs. \cite{Rost2006,FeyTrimer} found that the eigenvalues of Eq. \ref{triloeq} are essentially identical to those computed using the full Rydberg basis, consisting of states with finite quantum defects and multiple Rydberg manifolds. This is because the coupling between the trilobite and these other states is negligible. However, this same treatment fails catastrophically for studies of butterfly states as it is impossible to obtain even qualitatively accurate predictions within perturbation theory. The $p$-wave shape resonance causes the scattering volume to diverge, and the butterfly potential surfaces are only constrained to finite values via coupling to additional Rydberg manifolds. Additionally, the butterfly potential surface plunges through and couples to all quantum defect states before being repelled from the lower Rydberg manifold \cite{HamiltonGreeneSadeghpour}. This coupling is quantitatively important.

These problems are addressed by including trilobite and butterfly dimer orbitals for several $n$ manifolds ($\mathcal{M}$ is the number of manifolds) as well as the atomic basis states which have non-zero quantum defects in a hybrid basis \cite{JPBdens,EilesTutorial}.  The potential energy surfaces obtained with this method are identical to those computed via the full diagonalization using the Rydberg basis implied in Eq. \ref{ham}, but still have the advantage of a much more compact matrix representation of size $\mathcal{M} (4N+(l_\text{min}+1)^2)$ rather than $\mathcal{M}n^2$. The Hamiltonian in this basis is written
\be
\widetilde{H}=\begin{pmatrix}O_{PP'} & 0 \\ 0 & 1_{QQ'}\end{pmatrix}^{-1}\begin{pmatrix}H_{PP'} &H_{PQ'}\\H_{QP'} & H_{QQ'}\end{pmatrix},
\ee where the sub-block $H_{PP'}$ of dimension $4\mathcal{M}N$, the quantum defect sub-block $H_{QQ'}$ of dimension $\mathcal{M}(l_\text{min}+1)^2$, and the overlap matrix $O_{PP'}$ have matrix elements
\begin{align}
\label{matelementpolymer1}
H_{PP'}&=-\frac{1}{2n^2}\+{\Upsilon_{pq,n}^{\alpha\beta}}\delta_{nn'}\\&\nonumber+ 2\pi \sum_{i=1}^N\sum_{\xi = 1}^4a_i^{(\xi)}\+{\Upsilon_{pi,n}^{\alpha\xi}}\+{\Upsilon_{iq,n'}^{\xi\beta}}\\
H_{QQ'}&= -\frac{\delta_{nn'}\delta_{ll'}}{2(n-\mu_l)^2} \\&+ 2\pi  \sum_{i=1}^N\sum_{\xi = 1}^4a_i^{(\xi)}\phi_{nlm}^\xi(R_i)^*\phi_{n'l'm'}^\xi(R_i)\nonumber\\
O_{PP'}&=\+{\Upsilon_{pq,n}^{\alpha\beta}}\delta_{nn'}.
\end{align}
Additionally, there are coupling terms between dimer orbitals and the low$-l$ quantum defect states,
\begin{equation}
H_{PQ'}= 2\pi  \sum_{i=1}^N\sum_{\xi = 1}^4a_i^{(\xi)}\+{\Upsilon_{p i,n}^{\alpha \xi}}\phi_{n'l'm'}^\xi(R_i).
\end{equation}
In our present calculations we use the $n=29,30$, and $31$ Rydberg manifolds and include quantum defects for $s$, $p$, and $d$ waves ($l_\text{min}=2$). These parameters give adequately converged potential energy surfaces.

\section{Analysis of adiabatic potential energy surfaces}
\label{sec:potentials}

 \begin{figure}[t]
\includegraphics[width = 0.48\textwidth]{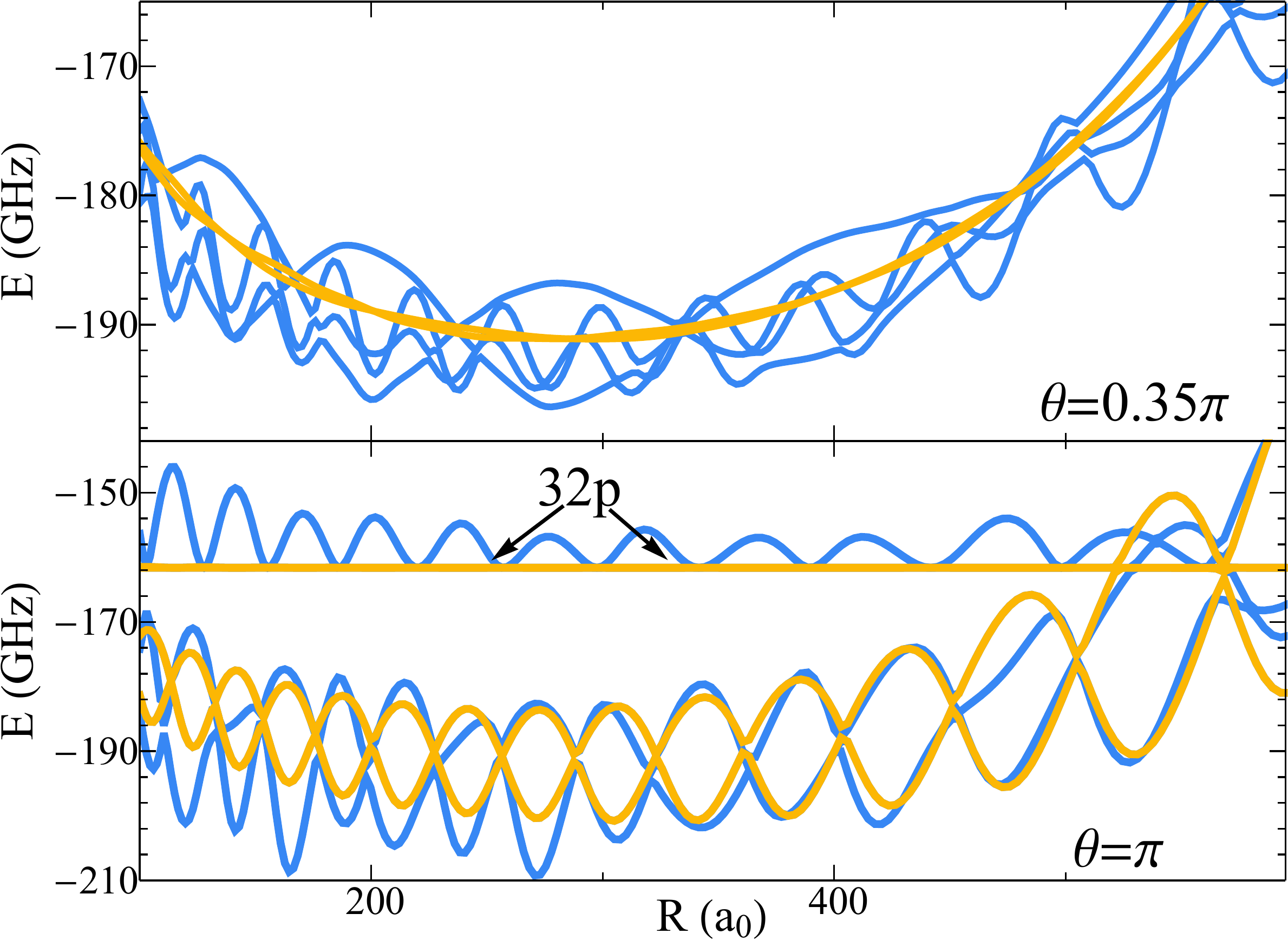}
\caption{Breathing mode slices of the Rb$_3$ potential surface for different bending angles $\theta$. Note the different energy axis in the two panels. The orange and blue curves have odd and even parity, respectively.  The energies are relative to the hydrogenic $n = 30$ energy. The potential energy surface associated with the $32p$ state cuts through the lower panel.} 
 \label{breathingmode}
\end{figure}

 For the paradigmatic Rydberg state studied here, $n=30$, the allowed range of internuclear distances for the butterfly states is $ R_i\in(100, 600)$. This range is determined by the energy dependence of the $p$-wave scattering volume, which in turn depends on $R$ through the semiclassical kinetic energy of the electron, $k(R)^2 = 2R^{-1}-n^{-2}$. The bond length $R_\text{res}$ associated with the shape resonance is therefore determined by the equation $2R_\text{res}^{-1}-n^{-2} = 2E_\text{res}$.  For $n=30$, $R_\text{res}\approx 600\Bohr$. It varies slowly as a function of $n$. The typical bond lengths of these molecules therefore do not scale proportional to $n^2$ as in the trilobite molecules. Just as $E_\text{res}$ varies among atomic species, so do these internuclear distances \cite{EilesHetero}.

\begin{figure}[t]
\includegraphics[width =1\columnwidth]{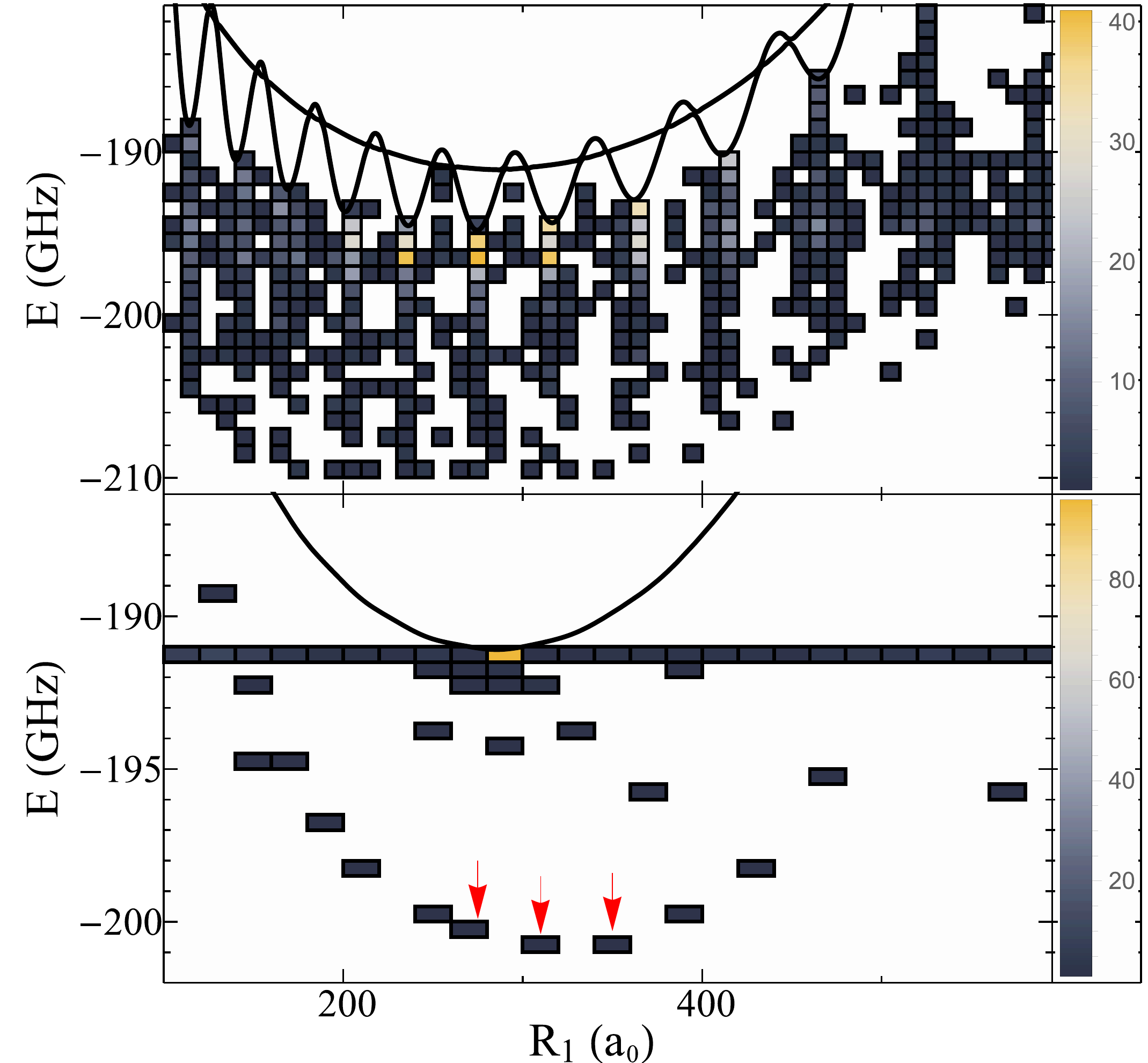}
\caption{Analysis of the energies and positions of trimer minima. Top: even-trimer; bottom: odd-trimer. The color code indicates the number of minima per bin at energy $E$ and position $R_1$. The dimer potential energy curves are shown in black. States considered in Sec. \ref{sec:states} are highlighted with red arrows. } 
 \label{resmin1}
\end{figure}

The three-dimensional potential surfaces 
 are challenging to visually investigate. Therefore, to get an impression of their behavior, we first exhibit in Fig. \ref{breathingmode} the breathing mode potential curves (symmetric stretch vibration): these are cuts through the surface at varying $R = R_1 = R_2$ and fixed $\theta$.   The two odd trimer potential curves (orange) oscillate dramatically in the collinear configuration, but they are very smooth and almost degenerate at  $\theta=0.35\pi$.  This is generally true for most angles $\theta/\pi <1$, where the odd trimer curves are nearly identical to the diatomic potential energy curve for the $\xi=4$ state (see Fig. \ref{resmin1}).  The four even trimer curves, on the other hand, oscillate for all values of $\theta$. When $\theta = \pi$, two of the potential curves become degenerate with the odd-trimer curves. For $\theta\ne\pi$, the even trimer curves separate into two sets having relatively fast and slow oscillation frequencies, respectively. The quickly (slowly) varying curves are predominantly mixtures of $R_1$ and $R_2$ ($\theta_1$ and $\theta_2$) dimer orbitals, as the coupling $\Upsilon_{12}^{23}$ is typically small.

\begin{figure}[t]
\includegraphics[width = 0.6\textwidth]{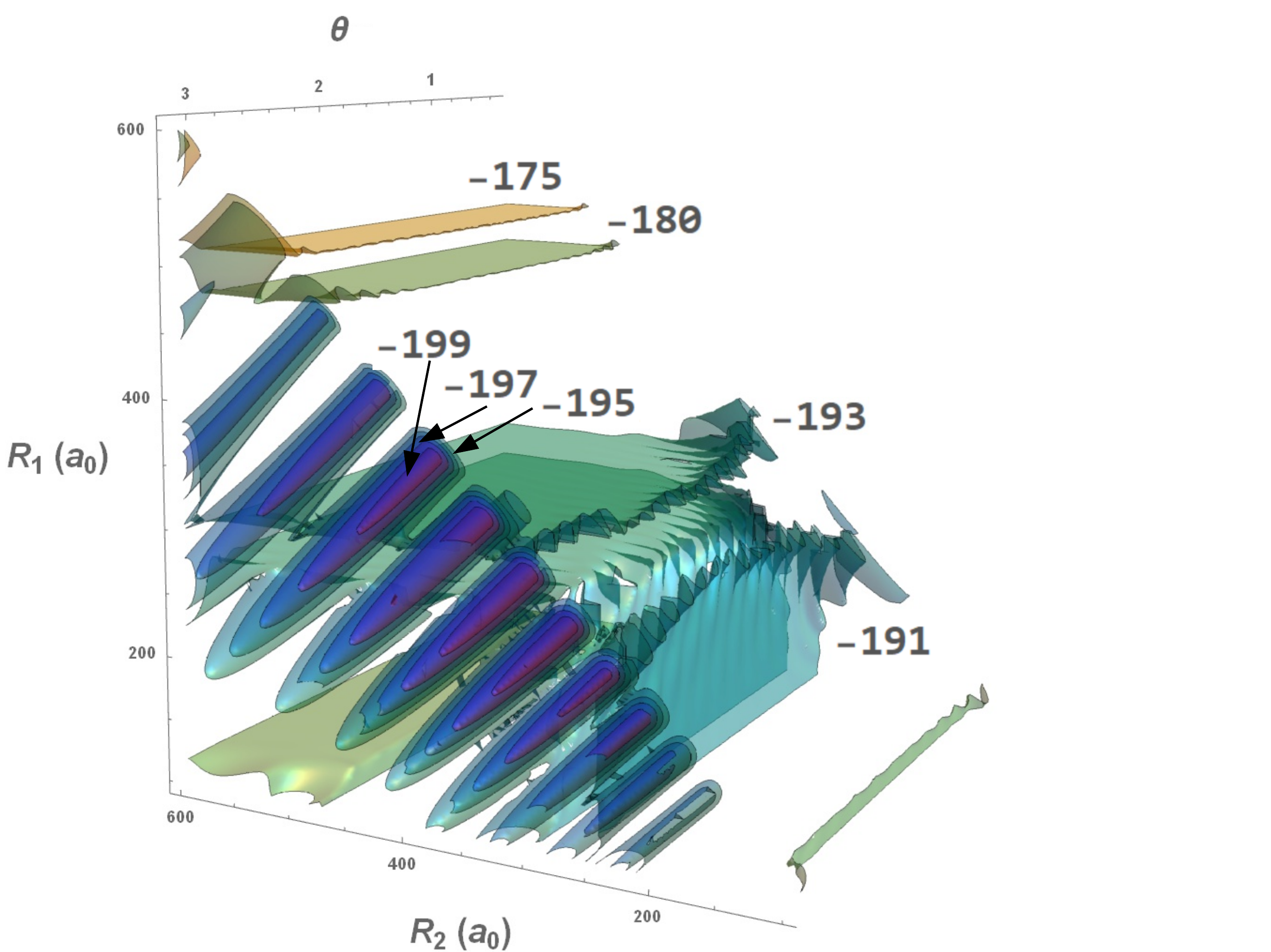}
\caption{Contours of the complete potential energy surface for the odd trimer state. Each contour, specified by a different colour, represents a surface of constant energy whose value is given in GHz in each label. The deepest contours are only found near the $\theta = \pi$ plane, and the higher energy contours are insensitive to changes in $\theta$. Since the potentials are symmetric with respect to reflection across the $R_1 = R_2$ line, the contours are not shown for $R_1>R_2$ for clarity.  } 
 \label{surfaces1}
\end{figure}

\begin{figure}[t]
\includegraphics[width = 0.50\textwidth]{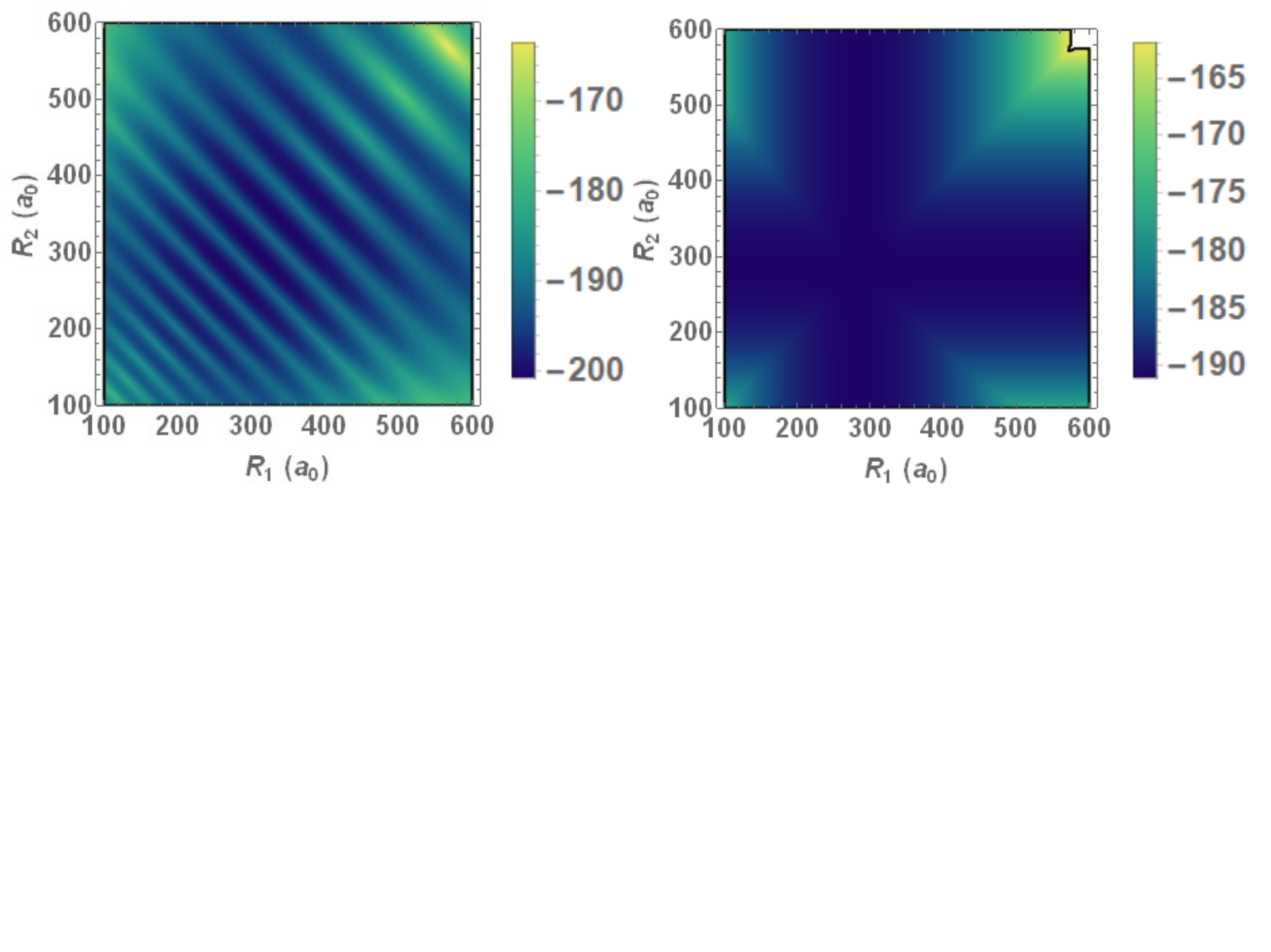}
\vspace{-100pt}
\caption{Odd parity potential surfaces at $\theta=\pi$ and $\theta=0.35\pi$ (left and right, respectively). The units of potential energy are in GHz. } 
 \label{fig:phipot}
\end{figure}

To gain further insight, we must move beyond the impediment of these low-dimensional potential cuts. Fig. \ref{surfaces1} presents the energetically deepest potential energy surface of the odd trimer state as a contour plot in the three nuclear coordinates. The deepest energy contours are only found near the collinear geometry ($\theta = \pi$), where cylindrically shaped wells around deep minima can be found. As $\theta$ decreases from $\pi$, the potential surface quickly become independent of $\theta$, and hence the constant energy contours become quite flat.
Fig. \ref{fig:phipot} shows the full potential energy surface $V(R_1,R_2,\theta)$ for the same two fixed angles $\theta = \pi,0.35\pi$ as in Fig. \ref{breathingmode}. This shows in more detail the deep potential wells in the collinear geometry and the nearly featureless surfaces away from this equilibrium position. 

These properties of the potential surfaces can be understood by analyzing the qualitative structure of the Hamiltonian more closely, focusing on the $\xi=4$ sub-block of Eq. \ref{matrix}.  In this subspace the two odd trimer potential surfaces are
 \begin{align}
\varepsilon_\pm(\vec R_1,\vec R_2) &= \frac{\varepsilon_d(R_1)+\varepsilon_d(R_2)}{2} \\&\pm \frac{1}{2}\sqrt{\left[\varepsilon_d(R_1) - \varepsilon_d(R_2)\right]^2+4c(\vec R_1,\vec R_2)},\nonumber
\end{align}where the cross term is
\begin{align}
c(\vec R_1,\vec R_2) &=4a_1^{(4)}a_2^{(4)}\left|\sum_{l>l_\text{min}}^{n-1}D_{l1}(R_1,0)D_{l1}(R_2,\theta_{12})\right|^2;\\
D_{lm}(R,\theta) &= m\frac{u_{nl}(R)}{R^2}\frac{Y_{lm}(\theta,0)}{\sin\theta},\nonumber
\end{align}
and the dimer potential is
\be
\varepsilon_d(R) =a_1^{(4)}\sum_{l>l_\text{min}}^{n-1}\left|\frac{u_{nl}(R)}{R^2}\right|^2\frac{(2l+1)(l+1)l}{8\pi}. 
\ee

\noindent When $c(\vec R_1,\vec R_2)$ vanishes, the trimer potentials reduce to independent dimer potentials, $\varepsilon_+(\vec R_1,\vec R_2) =\varepsilon_d(R_1)$, $\varepsilon_-(\vec R_1,\vec R_2) = \varepsilon_d(R_2)$, which are smooth and have a single global minimum. The cross-term induces mixing and creates additional wells in the potentials; however, it depends very sharply on $\theta_{12}$, $c(\vec R_1,\vec R_2)\sim\frac{1}{\sin^4\theta_{12}}$. This explains the appearance of interesting triatomic features only near the collinear geometry where $\theta_{12} = \pi$.

 \begin{figure}[t]
\includegraphics[width = 0.5\textwidth]{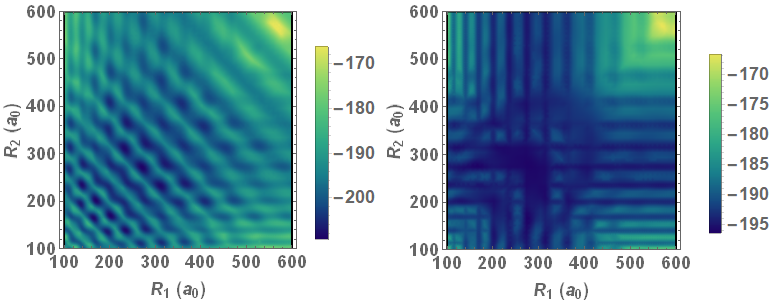}
\caption{Even trimer potential surfaces at $\theta=\pi$ and $\theta=0.35\pi$ (left and right, respectively). The units of potential energy are in GHz. } 
 \label{surfaces2}
\end{figure}

In contrast to the simplicity of the odd trimer potential surfaces, the even trimer potential surfaces are highly complex, varying much more rapidly as a function of all coordinates, but particularly as a function of $\theta_{12}$. For this reason a contour plot visualization is unintelligible, and we only show radial potential cuts in Fig. \ref{surfaces2} at the same two angles as before. These have minima at a plethora of, and hence flexibility in, stable molecular geometries. By comparing the two panels in Fig. \ref{surfaces2} we find that the potential wells in which these minima form are much more isolated from one another in the collinear geometry, and hence more suitable for localizing nuclear wave functions. 

\begin{figure*}[t]
\includegraphics[width = \textwidth]{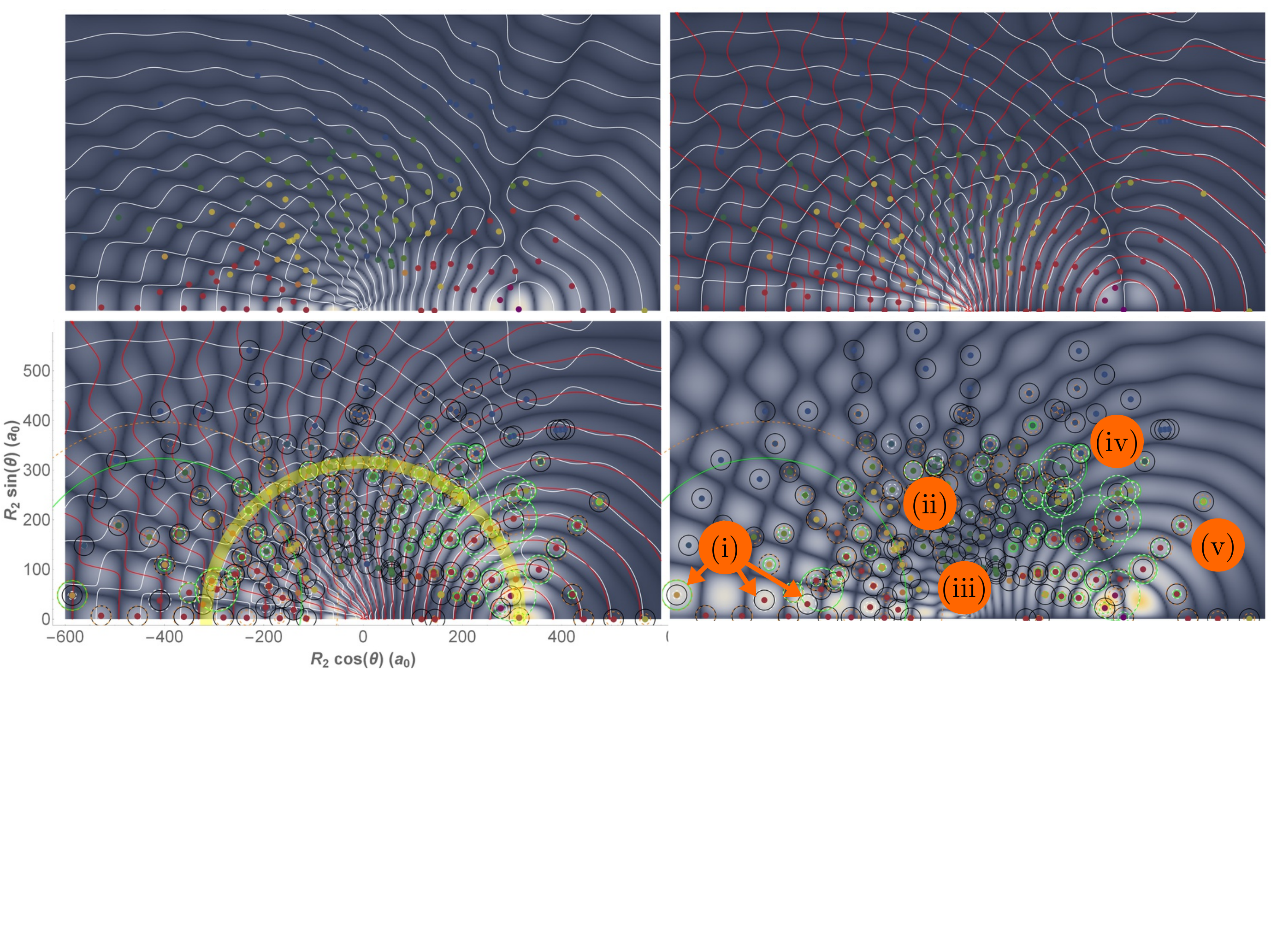}
\vspace{-125pt}
\caption{A study of the even-trimer with one bond length fixed at $R_1 = 316 \Bohr$. A different density plot is shown in each panel: $\Upsilon_{12}^{33}$ in the top left, $\Upsilon_{12}^{32}$ in the top right; $\Upsilon_{12}^{22}$ in the bottom left, and $\Upsilon_{12}^{23}$ in the bottom right. The density is largest when the colour shade is yellow, ranging through shades of white as it decreases, and finally dark gray when the density vanishes.   Positions of the potential energy minima are shown as colored dots: purple for -200 GHz, red for -198 GHz, yellow for -196 GHz, green for -194 GHz, and blue for -192 GHz.  The contour lines plotted correspond to $\Upsilon_{12}^{21}=0$ (red) and $\Upsilon_{12}^{31}=0$ (white). The yellow circle indicates $R_2 = R_1$, and the remaining circles provide information about the electronic state. Their radii indicate the  amplitude of each dimer orbital, normalized to the $R_1$-butterfly orbital (black). The $R_2$-butterfly is orange; the $\theta_2$-butterfly is white, and the $\theta_1$-butterfly is green.   Regions of interest expounded upon in the text are labeled in orange.  
}
 \label{minimapos1}
\end{figure*}
These analyses of the potential energy surfaces showed that for both types of trimers the collinear geometry is preferred.  We now turn to the question of which bond lengths are optimal. After finding minima in the complete potential surface, we bin them as a function of $R_1$ and energy and show them alongside the dimer potential curves in Fig. \ref{resmin1}. The even trimer minima are clustered around and typically just slightly deeper in energy than the dimer minima, but a sizable number of trimer minima are found as far as 20~GHz deeper in energy. From this one can conclude that, although there are exceptions, many trimer minima are found when one bond length is equal to a stable dimer bond length. In the following section we will analyze the specific case $R_1 = 316\Bohr$ to study the dependence of the minima depths on  $R_2$ and $\theta$, and thus characterize a subset of the stable geometries.  

In contrast, the histogram of odd-trimer minima is strongly peaked at the trivial global minimum, $R_0\approx 286\Bohr$, of the dimer potential. At a given $R_1 \ne R_0$ the minimum is most likely found at $R_2 = R_0$ unless the coupling term is very large, which explains the flat band of energies at the dimer minimum, around $-191$ GHz. At the collinear geometry the coupling becomes large, leading to a few non-trivial deep minima in the semicircle band. The global minimum shifts to $R_1\approx 316$. We will investigate the vibrational states associated with the marked bond lengths in Sec. \ref{sec:states}.

\section{Building principles for the even trimer}
\label{sec:minima}

For the pure trilobite sector of the triatomic Rydberg molecule, Eq.~\ref{triloeq}, it is clear that trimer minima are only found when the coupling element, $\Upsilon_{12}^{11}$, is large. This element, by definition, is the electronic trilobite wave function associated with a perturber at $\vec R_1$, evaluated at $\vec r = \vec R_2$; the trimer minima therefore occur when one perturber sits in a local maximum of the electronic wave function of a stable trilobite dimer. In the butterfly case, as suggested by Eq.~\ref{matrix} and resulting from the vectorial nature of the $p$-wave operator, the situation is more complex. There are now four possible coupling elements to maximize:  $\Upsilon_{12}^{22}$ and $\Upsilon_{12}^{32}$, which correspond to the \textit{derivative} $\frac{\partial }{\partial R}$ of the $R$-butterfly and the $\theta$-butterfly, and $\Upsilon_{12}^{33}$ and $\Upsilon_{12}^{23}$, the derivative $\frac{\partial }{\partial\theta}$ of the same respective orbitals. Rather than placing the second perturber at a maximum of the dimer orbital, a stable butterfly trimer is probable when the perturber is placed at point of locally steepest ascent or descent of the wave function of a stable dimer. 

As such, in  Fig.~\ref{minimapos1} we show density plots of the four derivative terms discussed above, and by overlaying the minima found at $R_1 = 316\Bohr$ we can correlate the potential energy minima positions to the orbital gradient maxima. Most commonly, we find that the minima lie at a maximum of at least one of these gradients, which also typically coincide with a node of the wave function, as depicted by the white ($\xi=2$) and red ($\xi=3$) contours. This relationship is useful for understanding the proliferation of potential minima and the types of electronic states and the couplings between the two butterfly orbitals allowed.

Many of these potential minima lie on several gradient maxima; this leads to a rich diversity of coupling strengths and mixing of the electronic states, as depicted using circles around each minimum in the bottom panels of Fig.~\ref{minimapos1}. Each circle radius equals the amplitude of each butterfly orbital, normalized to the $R_1$-butterfly amplitude. Along the curve $R_1 = R_2$, shown in yellow, the electronic state is typically an equal mixture of $R_1$- and $R_2$-butterflies. Deep within this circle and far outside of it, the electronic states tend to be dominated  by the $R_1$-butterfly component. Many of the remaining states are mixtures of either the $R_i$-butterflies or $\theta_i$-butterflies.  These, along with some interesting exceptions to this rule, are clustered into regions of the phase space labeled on the figure:
\begin{enumerate}
\item The states indicated by the arrows near this marker are unique in that they are mixtures of $\xi=2$ and $\xi=3$ orbitals. To the immediate left and right of the marker lie states dominated by $\theta_1$- and $R_2$- butterflies (green and orange, respectively). Further to the right is the opposite case, an equal mixture of $R_1$- and $\theta_2$- butterflies (black and white, respectively) with vanishing $R_2$ and $\theta_1$ components. These potential minima are located at nodes of the $\Upsilon_{12}^{22}$ and $\Upsilon_{12}^{33}$ derivative surfaces, but at maxima in the cross-term surfaces. A trimer of this type can have unusual electronic properties as a result of this coupling, since the $\xi=2$ and $\xi=3$ dipole moments have very different magnitudes and even different signs \cite{KhuskivadzePRA,EilesSpin}.
\item Immediately above this marker are two states with nearly equal contributions from all four dimer orbitals; along with a similar state at $(x,y) \approx (-380,60)$, these seem to be the only states with this composition for $R_1 = 316\Bohr$. 
\item Along the $y=0$ line, the electronic states change from primarily dimer-like at small $R_2$ to very trimer-like as $R_2$ increases. By dimer-like we mean that the eigenstate is dominated by the butterfly orbital of a single ground state atom, whereas trimer-like refers to a state which has significant contributions from butterfly states for each orbital, for example an even mixture of $R_1$- and $R_2$- butterflies. These potential minima are the deepest found in this geometry and are better isolated from nearby minima than most of the other configurations, allowing for tight localization of vibrational states. 
\item In this cluster, the states have large $\theta_{1}$- and $\theta_2$- butterfly mixing (green and white circles), in nearly equal amounts. Further to the left, closer to the (ii) marker, the opposite is mostly true (black and orange circles). 
\item Along this ray, the mixing between all orbitals is high and the minima are deep and well-isolated from other minima out to quite large distances. 
\end{enumerate}
Our goal in this phenomenological description of the electronic and nuclear state-space is not to systematically describe or classify the trimer states, but simply to reveal some of the rich diversity of possible electronic configurations. 

\section{Nuclear wave functions and vibrational spectra}
\label{sec:states}

\begin{figure*}[t]
\includegraphics[width = 0.99 \textwidth]{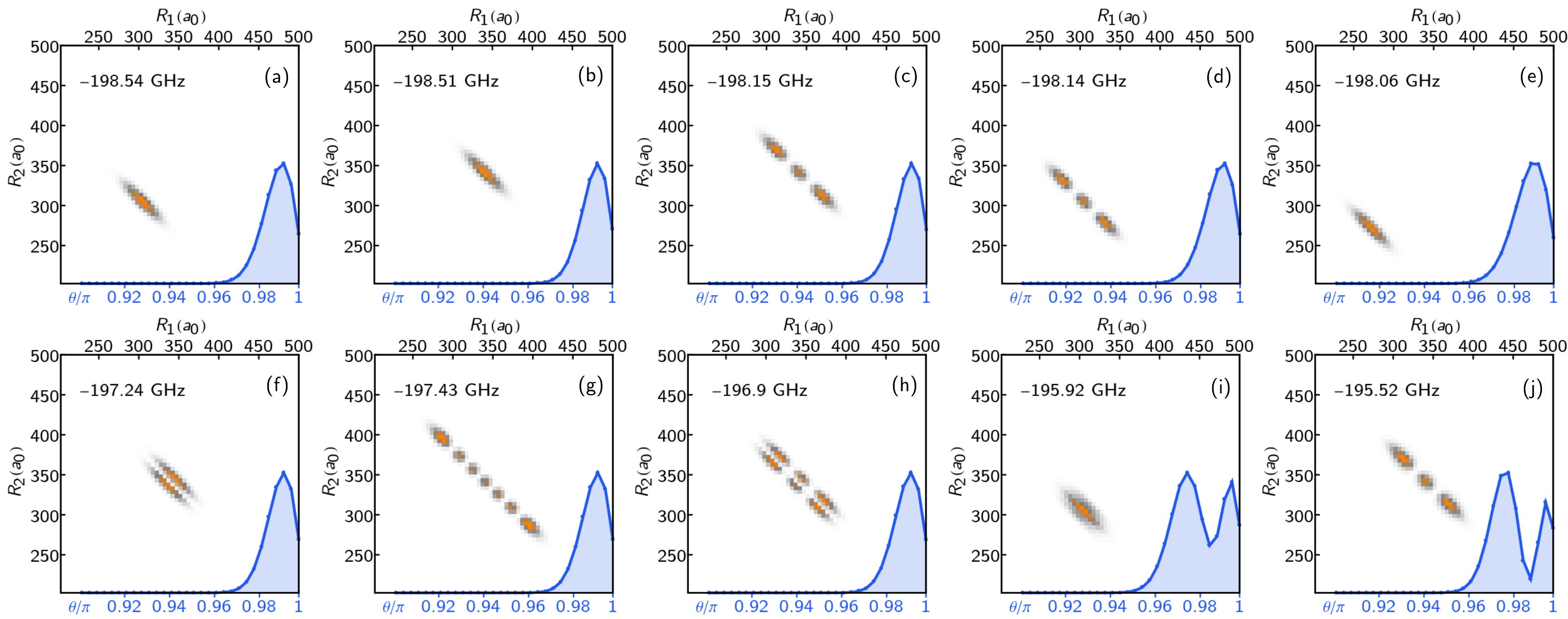}
\caption{Vibrational states of the odd trimer. Reduced radial probability densities,  $\int |\chi(r_1,r_2,\theta)|^2 \sin(\theta) \dd\theta$, are presented for different states together with their reduced angular densities, $\int |\chi(r_1,r_2,\theta)|^2 \sin(\theta) \dd r_1 \dd r_2$, (orange and blue, respectively) and are labeled by their vibrational energies. (a)-(e) are the five energetically lowest states in the specified coordinate range while (f)-(j) are a selection of excited states that illustrate excitations of additional bending and stretching modes. (f) and (h) show symmetric stretch excitations of the configurations of (b) and (c), respectively, while (g) is a highly excited asymmetric stretch state of (b). (i) and (j) have the same stretching excitation as (a) and (c), but an additional bending excitation. }  
\label{phi_states}
\end{figure*}

Having discussed the structure and the arrangement of minima in the potential energy landscape, we can now present the properties of supported vibrational states.  
We obtain vibrational wave functions $\chi(R_1,R_2,\theta)$ numerically as eigenstates of the vibrational Hamiltonian
\begin{align}
&H^\text{vib}= \nonumber \\
&\frac{1}{m}\left[-\frac{\partial^2}{\partial R_1^2}-\frac{\partial^2}{\partial R_2^2}- \cos \theta \frac{\partial}{\partial R_1}\frac{\partial}{\partial R_2}  \right] \nonumber \\
&- \frac{1}{m} \left(\frac{1}{R_1^2}+ \frac{1}{R_2^2}- \frac{\cos\theta}{R_1 R_2} \right)\left(\frac{\partial^2}{\partial \theta^2} + \cot \theta \frac{\partial}{\partial \theta}  \right) \nonumber \\
&-\frac{1}{m}\left(\frac{1}{R_1 R_2}-\frac{1}{R_2}\frac{\partial}{\partial R_1}
-\frac{1}{R_1}\frac{\partial}{ \partial R_2}
\right)\left(\cos \theta + \sin \theta \frac{\partial}{\partial \theta} \right) \nonumber\\
&+ \epsilon(R_1,R_2,\theta) \ ,
\label{eqn:Hamiltonian_nuclear}
\end{align}
where $m$ is the mass of $^{87}$Rb. 
This Hamiltonian describes the pure vibrational dynamics of the trimer (depending only on $R_1$, $R_2$, $\theta$) and can be obtained from the full nuclear Hamiltonian by separating the center-of-mass motion and projecting onto the subspace of conserved relative angular momentum $L=0$ \cite{Carter1982, Handy1987, FeyKurz, FeyTrimer}.  The wave functions are normalized as $\int \dd R_1 \dd R_2 \dd \theta \sin \theta |\chi(R_1,R_2,\theta)|^2=1$. In our numerical approach we construct the Hamiltonian on a three-dimensional grid in position space using a finite difference representation for the radial degrees of freedom, $R_1$ and $R_2$, and a discrete variable representation for the $\theta$ direction \cite{Beck2000}. According to the spin statistics of $^{87}$Rb, we consider only bosonic states with $\chi(R_1,R_2,\theta)=\chi(R_2,R_1,\theta)$.

\begin{figure*}
\includegraphics[width = 0.99 \textwidth]{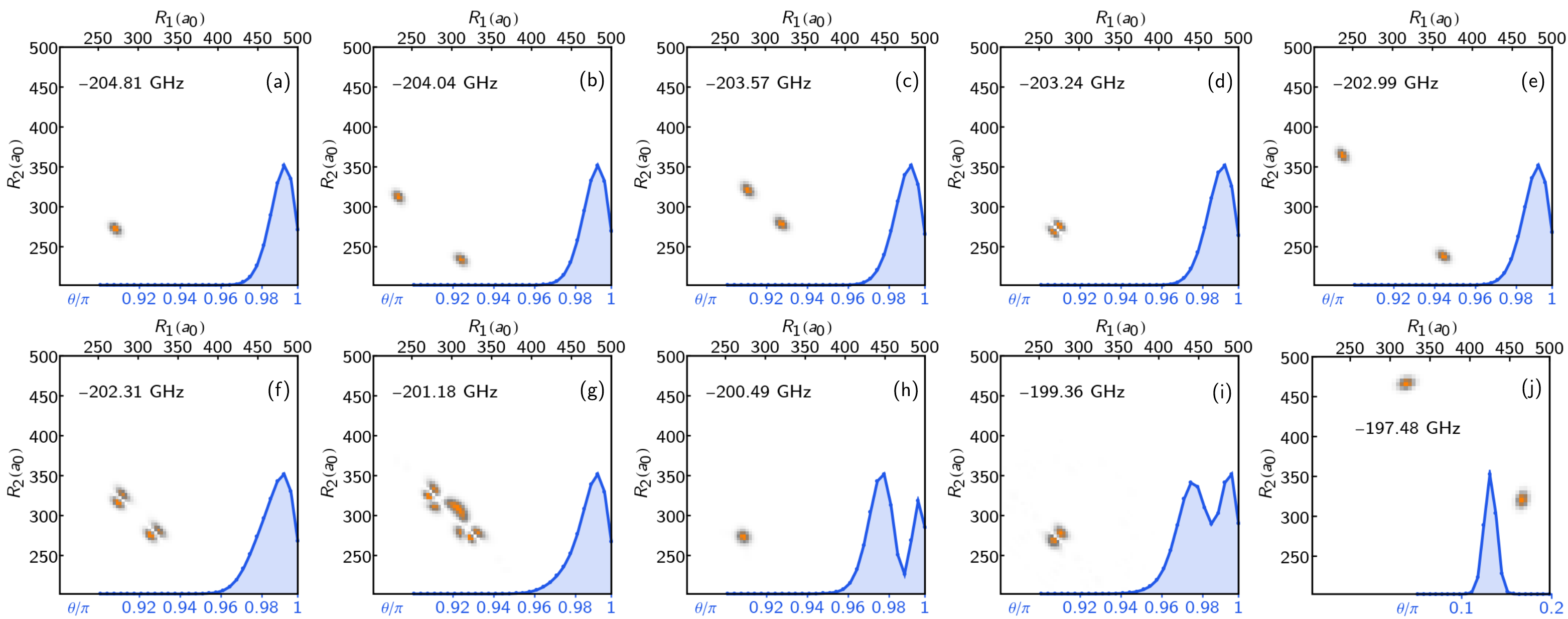}
\caption{Vibrational states of the even trimer with similar quantities as in Fig. \ref{phi_states}. (a)-(e) are the five energetically lowest states in the specified coordinate range while (f)-(j) are selected excited states that illustrate effects of delocalization (g) as well as excitations of additional bending (h), (i) and stretching (f), (i) modes, as well as a non-collinear configuration (j).}  
\label{r_states}
\end{figure*}

Fig. \ref{phi_states} (a)-(e) presents energies and reduced probability densities of the energetically lowest vibrational states of the odd butterfly.
The reduced densities are obtained by averaging the density $|\chi(R_1,R_2,\theta)|^2$ over one or two degrees of freedom, respectively, and contain information about the likelihood to find the trimer in a certain geometry. 
For instance, we can infer from the radial densities (orange) that the ground-state (a) has a bond length of $R_1=R_2=316\Bohr$. This state is 30 MHz detuned from the next excited state (b) with a bond length of $R_1=R_2=340 \Bohr$. Both states are spatially well separated and strongly confined to a collinear geometry as indicated by the angular density (blue). The following states (c) and (d) correspond to excitations of asymmetric stretching modes of the states (b) and (a), respectively. The bosonic spin statistics are reflected here by the absence of states with odd numbers of nodes along the asymmetric stretch mode. State (e) populates yet another equilibrium geometry with a bond length of  $R_1=R_2=270\Bohr$. A selection of even higher lying states is shown in Fig. \ref{phi_states} (f)-(j).  Some of them exhibit excitations of symmetric stretching modes (f),(h), highly excited asymmetric stretching modes (g), as well as bending modes (i),(j).    
To an excellent approximation states belonging to the same bond length are well described by an harmonic ladder with spacings of  approximately 400 MHz for the asymmetric stretching, 1.2 GHz for the symmetric stretching, and 2.5 GHz for the bending modes. Importantly, the order of these spacings differs from triatomic low-$l$ Rydberg molecules, where the energy spacing of bending motion is typically much smaller than stretching motion \cite{FeyKurz}. It also differs from the trilobite trimer, as shown by an example in reference \cite{FeyTrimer} where the spacing of bending and stretching modes is almost equal.

Reduced probability densities of the even butterfly trimer are shown in Fig. \ref{r_states}. Again we present the energetically lowest states (a)-(f) as well as a selection of excited states. Due to the different potential energy surface, the vibrational states of the even trimer differ accordingly in some of their properties from the odd trimer states. First, there are not only symmetric states where both ground-state atoms share the same bond length (as in panel (a)) but also states where one ground-state atom is always closer to the Rydberg core than the other one, see e.g.~(b), (c) and (e). However, due to the bosonic character of the ground-state atoms all densities are symmetric under reflection with respect to the $R_1=R_2$ diagonal. When these states become excited they can couple to states belonging to different equilibrium configurations to form more complex superpositions, e.g.~(g). A second difference is the possibility to form well-localized states in non-colinear arrangements (j). Despite their much more complex underlying electronic structure, vibrational states of the even butterfly trimers are in this respect very similiar to trilobite trimers \cite{FeyTrimer}.

\section{Conclusions}
\label{sec:conclusions}
We have extended the analysis of triatomic ultra-long-range Rydberg molecules begun in Refs. \cite{FeyKurz,FeyTrimer}, which focused on low-$l$ states and trilobite states, to the butterfly states. These trimers come in two varieties which behave very differently. The odd trimers are only stable in the collinear geometry, have very few equilibrium geometries with equal bond length $R_1 = R_2$, and due to the simple structure of the potential surface at $\theta=\pi$ have nearly independent asymmetric stretch, symmetric stretch, and bending modes.  In contrast, the even trimers exhibit complex and vibrant potential surfaces with a rich pattern of potential wells; many of these, however, are insufficiently isolated from other wells to localize vibrational states. As in the odd trimers, many -- but not all -- of the even trimers have a collinear geometry, but many have equilibrium positions at $R_1 \ne R_2$ and have a more complex excitation spectrum.

Although we have not mentioned in detail the large dipole moments of the dimer orbitals, many of the equilibrium configurations of the even trimer will have non-zero dipole moments stemming from either the asymmetry in bond lengths or the mixing of $R$- and $\theta$- dimer orbitals. These trimers therefore possess interesting field control possibilities \cite{KurzSchmelcherPRA}. In the present study we neglected the complex spin structure of these molecules. Including these effects in a polyatomic context is challenging, but will be necessary for quantitative predictions \cite{FeyNew}. The mixing of symmetric and asymmetric stretch modes hinted by Fig. \ref{r_states}(g) implies that the dynamical behavior of vibrational wave packets across these oscillatory potential surfaces will likely be very rich and offer interesting avenues to explore non-adiabatic physics. 

\ack{M.T.E acknowledges support from the Max-Planck Gesellschaft via the MPI-PKS visitors program and from an Alexander von Humboldt Stiftung postdoctoral fellowship.  F.H. and P.S. acknowledge support from the Deutsche Forschungsgemeinschaft 
within the priority program "Giant interactions in Rydberg systems" [DFG 
SPP 1929 GiRyd project SCHM 885/30-1].  }
\section*{References}
\bibliographystyle{iopart-num}

\providecommand{\newblock}{}

\end{document}